\documentclass[aps,prd,amsmath,floats,floatfix,twocolumn,superscriptaddress,%
nofootinbib,showpacs]{revtex4}
\usepackage{graphicx,amssymb,amsmath,amsbsy,mathrsfs}
\usepackage{bm}

\usepackage[usenames]{color}

\newcommand{\Mirr}{M_{\text{irr}}}

\newcommand{\SMM}{\chi}

\newcommand{\SpinFromShapeMin}{\TypeSC{\SMM}^{\text{min}}}
\newcommand{\SpinFromShapeMax}{\TypeSC{\SMM}^{\text{max}}}

\newcommand{\TypeSC}[1]{#1_{\rm SC}}

\newcommand{\Caltech}{\affiliation{Theoretical Astrophysics 350-17,
    California Institute of Technology, Pasadena, CA 91125}}

\newcommand{\CITA}{\affiliation{Canadian Institute for Theoretical Astrophysics,
60 St. George Street, University of Toronto, Toronto, ON M5S 3H8, Canada}}

\begin{document}
\vspace{-2.5cm} 

\title{High accuracy simulations of black hole binaries: 
  spins anti-aligned with the orbital angular momentum}

\author{Tony Chu} \Caltech
\author{Harald P. Pfeiffer} \Caltech\CITA
\author{Mark A. Scheel} \Caltech
\date{\today}

\begin{abstract}
High-accuracy binary black hole simulations are presented for
  black holes with spins anti-aligned with the
  orbital angular momentum.  The particular case studied represents an
  equal-mass binary with spins of equal magnitude
  $S/m^2=0.43757\pm0.00001$.  
  The system has initial orbital eccentricity $\sim 4\times
  10^{-5}$, and is evolved through 10.6 orbits plus merger and ringdown.
  The remnant mass and spin are
  $M_f=(0.961109 \pm 0.000003)M$ and $S_f/{M_f}^2=0.54781 \pm
  0.00001$, respectively, where $M$ is the mass during early inspiral. 
The gravitational waveforms  
have accumulated 
numerical phase errors of $\lesssim$ 0.1 radians without
any time or phase shifts, 
and $\lesssim 0.01$ radians when the waveforms 
are aligned with suitable time and phase shifts.
The waveform is extrapolated to infinity 
using a procedure 
accurate to $\lesssim$ 0.01 radians in phase, and the extrapolated
waveform differs by up to 0.13 
radians in phase 
and about one percent in amplitude
from the waveform extracted at finite radius $r=350M$.  
The simulations employ different choices for the constraint damping 
parameters in the wave zone; this greatly reduces the effects of junk
radiation, 
allowing the extraction of
a clean gravitational wave signal even very early in the simulation.
\end{abstract}


\maketitle

\section{Introduction}

Much progress has been made in recent years in the numerical solution of 
Einstein's equations for the inspiral, merger, and ringdown of binary black 
hole systems. Since the work of Pretorius~\cite{Pretorius2005a} and the 
development of the moving puncture method~\cite{Campanelli2006a,Baker2006a}, 
numerical simulations have been used to analyze post-Newtonian 
approximations~\cite{Buonanno-Cook-Pretorius:2007,Baker2006d,Pan2007,%
Buonanno2007,Hannam2007,Boyle2007,Gopakumar:2007vh,Hannam2007c,DN2007b,%
Damour2007a,Boyle2008a,Mroue2008,Hinder2008b,DN2008,Damour2009a,%
Buonanno:2009qa}, to investigate the recoil velocity of the final black 
hole~\cite{Campanelli2005,Herrmann2007b,Baker2006c,Gonzalez2007,%
Campanelli2007,Gonzalez2007b,Bruegmann-Gonzalez-Hannam-etal:2007,%
Herrmann2007,Herrmann2007c,Choi-Kelly-Boggs-etal:2007,Baker2007,%
Tichy:2007hk,Schnittman2007,Campanelli2007a,Koppitz2007,MillerMatzner2008,%
Baker2008,Healy2008,Rezzolla:2007xa}, 
and to explore the the orbital dynamics of spinning 
binaries~\cite{Campanelli2006c,Campanelli2006d,Campanelli2007b,Herrmann2007c,%
MarronettiEtAl:2008,Berti2007b}.

Numerical simulations can provide an accurate knowledge of gravitational 
waveforms, which is needed to make full use of the information obtained from 
gravitational-wave detectors such as LIGO and LISA. Not only can detected 
gravitational waveforms be compared with numerical results to measure 
astrophysical properties of the sources of gravitational radiation, but the 
detection probability itself can be increased via the technique of matched 
filtering~\cite{Finn1992}, in which noisy data are convolved with numerical 
templates to enhance the signal.

The production of accurate numerical waveforms is computationally 
expensive, making it challenging to construct an adequate waveform template 
bank covering a sufficiently large region of the parameter space of black hole 
masses and spins. 
One way of increasing efficiency is to adopt techniques known as spectral
methods.  For smooth solutions, 
spatial discretization
errors of spectral methods decrease exponentially with increasing
numerical resolution. In contrast, errors decrease polynomially for the
finite difference methods used in most binary black hole simulations.
Not only have spectral methods been
used to prepare very accurate initial data~\cite{Bonazzola1996,Bonazzola1999a,%
grandclement-etal-2001,Gourgoulhon2001,Grandclement2002,Pfeiffer2002a,%
Pfeiffer2003,Cook2004,AnsorgBruegmann2004,Ansorg:2005,Caudill-etal:2006,%
Grandclement2006,Lovelace2008,FoucartEtAl:2008,Buchman:2009ew}, 
but they have been used 
to generate the longest and most accurate binary black hole simulation to 
date~\cite{Scheel2008}.

Following the previous work of~\cite{Scheel2008}, this paper presents
the first spectral simulation of an orbiting and merging binary with 
spinning black holes: an equal mass system with spins of 
the black holes anti-aligned with the orbital angular momentum.  Simulations of
binaries with spins parallel to the orbital momentum are certainly not
new, e.g.~\cite{Campanelli2006c,Herrmann2007c,Koppitz2007,%
Pollney-Reisswig:2007,MarronettiEtAl:2008,Rezzolla:2007xa,Hannam2007c}.
Our goal here is to show that such systems can be simulated with
spectral methods, and that the high accuracies achieved for the
non-spinning case carry over into this more general regime.

The spin of each black hole is $S/m^2=0.43757\pm 0.00001$.
The determination of this quantity, as well as other spin measures, 
is
explained in more detail in
Sec.~\ref{sec:BHSpinsMasses}. The
evolution consists of 10.6 orbits of inspiral with an orbital
eccentricity of $e \sim 4\times10^{-5}$, followed by the merger and
ringdown.  We find that this simulation has accuracy comparable 
to that of the simulation presented in~\cite{Scheel2008}.  
We also present different choices for the constraint damping
parameters in the wave zone;  
these choices cause the initial noise (``junk radiation'')
to damp more rapidly, resulting in a useable, almost noise-free waveform 
much earlier in the simulation.

This paper is organized as follows:
In Sec.~\ref{sec:ID}, we discuss the construction of our
initial data. In Sec.~\ref{sec:Evolutions}, we describe the equations,
gauge conditions, and numerical methods used to solve Einstein's
equations.  In Sec.~\ref{sec:NumericalProps}, we present several
properties of our simulations, including constraints, and the spins
and masses of the black holes. In Sec.~\ref{sec:Waveform}, we explain
the extraction of gravitational waveforms from the simulation, and the
extrapolation of the waveforms to infinity. Finally, in
Sec.~\ref{sec:Discussion}, we discuss outstanding difficulties and
directions for future work.

\section{Initial Data}
\label{sec:ID}

The initial data are almost identical to those used in the
  simulation of an equal-mass, non-spinning black hole binary
  presented in Refs.~\cite{Boyle2007,Scheel2008}.  We use
  quasi-equilibrium initial
  data~\cite{Cook2002,Cook2004,Caudill-etal:2006} (see
  also~\cite{Gourgoulhon2001,Grandclement2002}), built using 
 the conformal thin sandwich
  formalism~\cite{York1999,Pfeiffer2003b}, and employing the
  simplifying choices of conformal flatness and maximal slicing.
Quasi-equilibrium boundary conditions are imposed on spherical
excision boundaries for each black hole, with the lapse boundary
condition given by Eq.~(33a) of Ref.~\cite{Caudill-etal:2006}.  The
  excision spheres are centered at Cartesian coordinates
  $C_1^i=\left(d/2,0,0\right)$ and 
  $C_2^i=\left(-d/2,0,0\right)$, where we choose the same coordinate
  distance $d$ and the same excision radii as in~\cite{Boyle2007}.

Within this formalism, the spin of each black hole is determined by 
a parameter $\Omega_r$ and a conformal Killing vector $\xi^i$ 
(tangential to the excision sphere); these
enter into the boundary condition for the shift 
$\beta^i$ at an excision surface~\cite{Cook2004}.  We will
use the  sign convention of Eq.~(40) in Ref.~\cite{Lovelace2008}, 
so that positive $\Omega_r$ corresponds to corotating black holes.
The same value of $\Omega_r$ is chosen at both excision surfaces, 
resulting in black holes with equal spins.
In Refs.~\cite{Boyle2007,Scheel2008},
$\Omega_r$ was chosen to ensure 
that the black hole spins vanish~\cite{Caudill-etal:2006}.  
In this paper, 
we instead 
fix $\Omega_r$ at some 
negative value, resulting in moderately spinning black holes that
counterrotate with the orbital motion.

Two more parameters need to be chosen before initial data can be
constructed: The orbital angular frequency $\Omega_0$ and the radial
velocity $v_r$ of each black hole.  These parameters are determined
by an iterative procedure that minimizes the orbital eccentricity during
the subsequent evolution of the binary: We start by setting $\Omega_0$
and $v_r$ to their values in the non-spinning evolution of
Ref.~\cite{Pfeiffer-Brown-etal:2007}, we solve the initial value
equations a pseudo-spectral elliptic solver~\cite{Pfeiffer2003}, and
we evolve for about 1-2 orbits using the techniques described in
Sec.~\ref{sec:Evolutions}.  Analysis of this short evolution yields an
estimate for the orbital eccentricity, and improved parameters
$\Omega_0$ and $v_r$ that result in a smaller orbital eccentricity.
This procedure is identical to Ref.~\cite{Boyle2007}, except that we 
include a term quadratic in $t$ for the function used 
to fit the radial velocity ($ds/dt$),
to obtain better fits.
We repeat this procedure until the eccentricity of the black hole
binary is reduced to $e\sim 4\times10^{-5}$.  
Properties of this low-eccentricity initial data set are summarized in
the top portion of Table~\ref{tab:IDTable}.

\begin{table}
\begin{tabular}{|ll|}
\hline
{\bf Initial data} &\\ 
Coordinate separation & $d/M_{\rm ID}=13.354418$ \\
Radius of excision spheres & $r_{\rm exc}/M_{\rm ID}=0.382604$ \\
Orbital frequency & $\Omega_0 M_{\rm ID} = 0.0187862$ \\
Radial velocity & $v_r = -7.4710123\times 10^{-4}$ \\
Orbital frequency of horizons & $\Omega_r M_{\rm ID}= -0.242296$ \\ 
Black hole spins & $\chi_{\rm ID}=0.43785$ \\
ADM energy & $M_{\rm ADM}/M_{\rm ID} = 0.992351$\\
Total angular momentum & $J_{\rm ADM}/M_{\rm ID}^2 = 0.86501$ \\
Initial proper separation & $s_0/M_{\rm ID} = 16.408569$ \\ \hline
{\bf Evolution} &\\
Initial orbital eccentricity & $e\approx 4\times 10^{-5}$ \\
Mass after relaxation & $M=( 1.000273 \pm 0.000001)M_{\rm ID}$\\
Spins after relaxation & $\chi=0.43757 \pm 0.00001$\\
Time of merger (common AH) & $t_{\rm CAH}=2399.38 M$\\
Final mass  & $M_f=(0.961109 \pm 0.000003)M$\\
Final spin & $\chi_f= 0.54781 \pm 0.00001$
\\\hline 
\end{tabular}
\caption{
Summary of the simulation presented in this paper. 
The first block lists properties of the initial data, the second 
block lists properties of the evolution.
\label{tab:IDTable}}
\end{table}

The data in the upper part of Table~\ref{tab:IDTable} are
given in units of $M_{\rm ID}$,
the sum of the black hole masses in
the initial data.  For any black hole (initial data, during the evolution,
the remnant black hole after merger), we define its mass using
Christodoulou's formula,
\begin{equation}
\label{eq:Christodoulou}
m^2 = m_{\rm irr}^2 + \frac{S^2}{4m_{\rm irr}}.
\end{equation}
We use the apparent horizon area $A_{\rm AH}$ to define the
irreducible mass $m_{\rm irr}=\sqrt{A_{\rm AH}/(16 \pi)}$. 
The {\em nonnegative} spin $S$ of 
each black hole is computed with the spin diagnostics described
in~\cite{Lovelace2008}.  Unless noted otherwise, we compute the
spin from an angular momentum surface 
integral~\cite{BrownYork1993,Ashtekar-Krishnan:2004}
using approximate Killing vectors of the apparent horizons, as
described in~\cite{OwenThesis,Lovelace2008} (see
also~\cite{Dreyer2003,Cook2007}).  We define the dimensionless 
spin by 
\begin{equation}\label{eq:SMMDef}
\chi=\frac{S}{m^2}.
\end{equation}

\section{Evolutions}
\label{sec:Evolutions}

\subsection{Overview}

The Einstein evolution equations are solved with the pseudo-spectral
evolution code described in Ref.~\cite{Scheel2008}. 
This code evolves
a first-order representation~\cite{Lindblom2006} of the generalized
harmonic system~\cite{Friedrich1985,Garfinkle2002,Pretorius2005c}
and includes terms that damp away small constraint
violations~\cite{Gundlach2005,Pretorius2005c,Lindblom2006}.
The computational domain extends from excision boundaries located just
inside each apparent horizon to some large radius, and is divided into
subdomains with simple shapes (e.g. spherical shells, cubes, cylinders). 
No boundary conditions
are needed or imposed at the excision boundaries, because all characteristic
fields of the system are outgoing (into the black hole) there.  The boundary
conditions on the outer 
boundary~\cite{Lindblom2006,Rinne2006,Rinne2007} are designed to
prevent the influx of unphysical constraint
violations~\cite{Stewart1998,FriedrichNagy1999,Bardeen2002,Szilagyi2002,%
Calabrese2003,Szilagyi2003,Kidder2005} and undesired incoming gravitational 
radiation~\cite{Buchman2006,Buchman2007}, while allowing the outgoing 
gravitational radiation to pass freely through the boundary. Interdomain 
boundary conditions are enforced with a penalty
method~\cite{Gottlieb2001,Hesthaven2000}. 

The gauge freedom in the generalized harmonic formulation of Einstein's
equations is fixed
via a freely specifiable gauge source function $H_a$ that satisfies the
constraint
\begin{equation}
  \label{e:ghconstr}
  0 = \mathcal{C}_a \equiv \Gamma_{ab}{}^b + H_a,
\end{equation}
where $\Gamma^{a}{}_{bc}$ are the spacetime Christoffel symbols.
We choose $H_a$ differently during the inspiral, plunge and ringdown, 
as described in detail in Sections~\ref{sec:Inspiral},
\ref{sec:Plunge}, and~\ref{sec:Ringdown}.

In order to treat moving holes using a fixed grid, we employ multiple
coordinate frames~\cite{Scheel2006}: The equations
are solved in `inertial frame' that is asymptotically Minkowski, but
the grid is fixed in a `comoving frame' in which the black
holes do not move.  The motion of the holes is accounted for by
dynamically adjusting the coordinate mapping between the two 
frames\footnote{All coordinate quantities (e.g. trajectories, 
waveform extraction radii) in this paper
are given with respect to the inertial frame unless noted otherwise.}.
This coordinate mapping is chosen differently at different stages
of the evolution, as described in Sections~\ref{sec:Inspiral},
\ref{sec:Plunge}, and~\ref{sec:Ringdown}.

The simulations are performed at four different resolutions, N1 to N4.  
The approximate number of collocation points for these resolutions is given 
in Table~\ref{tab:RunTable}.

\begin{table}
\begin{tabular}{| c | c c c |}
\hline
Run & $N_{pts}$ & CPU-h & CPU-h/$T$ \\\hline
N1  &  $(64^3,65^3,65^3)$ & 9,930 & 3.4 \\\hline
N2  &  $(70^3,72^3,72^3)$ & 16,195 & 5.6 \\\hline
N3  &  $(76^3,78^3,80^3)$ & 28,017 & 9.7 \\\hline
N4  &  $(82^3,84^3,87^3)$ & 44,954 & 15.5 \\\hline
\end{tabular}
\caption{
Approximate number of collocation points and CPU usage for the 
evolutions performed here. The first column indicates the name 
of the run. $N_{pts}$ is the approximate number of collocation points used 
to cover the entire computational domain. The three values for $N_{pts}$ are 
those for the inspiral, plunge, 
and ringdown portions of the simulation, 
which are described in Sections~\ref{sec:Inspiral},~\ref{sec:Plunge} 
and~\ref{sec:Ringdown}, respectively. The total run times $T$
are in units of the total 
Christodoulou mass $M$ [cf. Eq.~(\ref{eq:Christodoulou})] of the binary.
\label{tab:RunTable}
}
\end{table}

\subsection{Relaxation of Initial Data}

\begin{figure}
\includegraphics[width=0.45\textwidth]{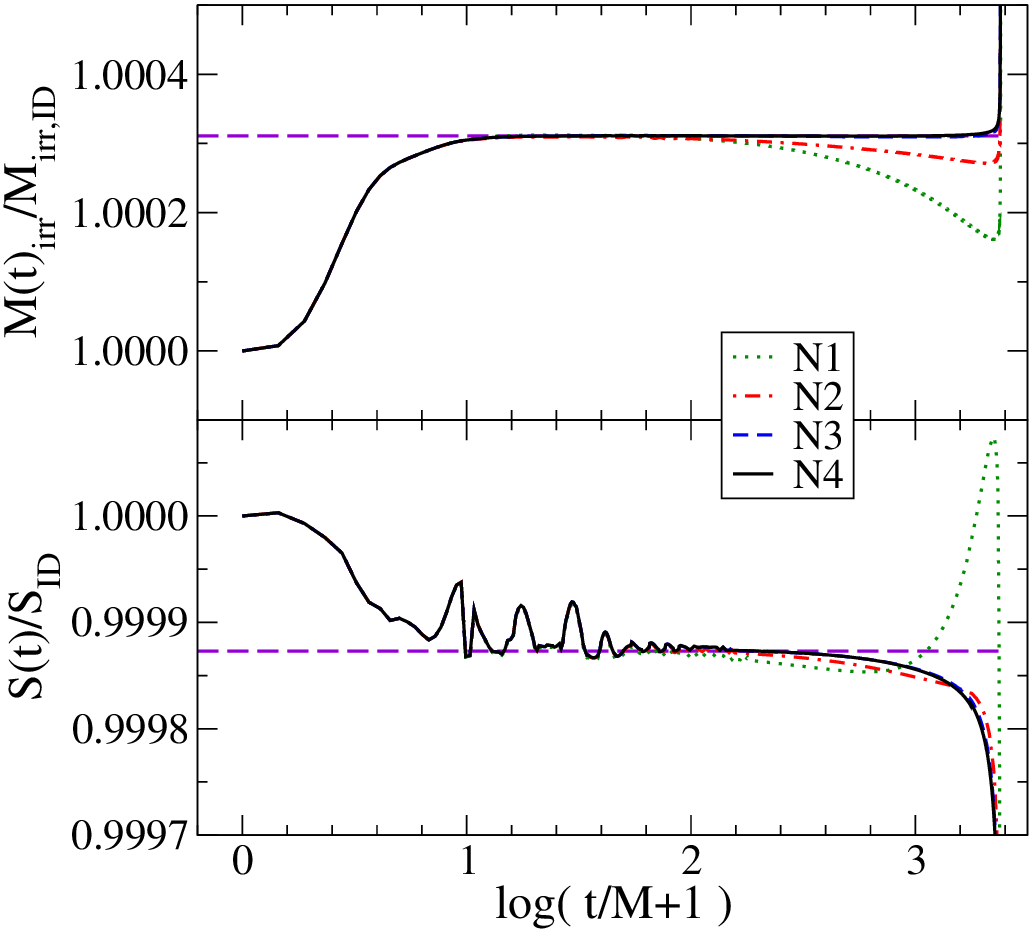}
\caption{\label{fig:TotalIrrMassAKVSpinForInspiral}
Irreducible mass (top panel) and spin (bottom panel) of the black
  holes during the relaxation of the initial data to the equilibriun
  (steady-state) inspiral configuration.  Shown are four different 
    numerical resolutions, N1 (lowest) to N4 (highest), 
    cf. Table~\ref{tab:RunTable}.  Up to $t\sim 10 M$, both
  mass and spin change by a few parts in $10^4$, then they remain
  approximately constant (as indicated by the
  dashed horizontal lines)
  until shortly before merger.  
  These steady-state values are used to define $M$ and $\chi$.
}
\end{figure}

The initial data do not precisely correspond to two black holes in
equilibrium, e.g., because tidal deformations are not incorporated
correctly, and because of
the simplifying choice of conformal flatness.
Therefore, early in the evolution the system relaxes
and settles down into a new steady-state configuration.
Figure~\ref{fig:TotalIrrMassAKVSpinForInspiral} shows the change in
irreducible mass and spin relative to the initial data during the
evolution.  During the first $\sim 10M$ of the evolution, $\Mirr$
increases by about $3$ parts in $10^4$ while the spin decreases by
about 1 part in $10^4$.  These changes are 
resolved  by
all four numerical resolutions, labeled N1 (lowest) 
to N4 (highest), and converge with increasing resolution. 
After the initial relaxation, for $10M\lesssim t \lesssim 2350M$, 
the mass is constant to about 1 part in $10^6$, as can be seen
from the convergence of the different resolutions in the upper panel
of Fig.~\ref{fig:TotalIrrMassAKVSpinForInspiral}.  In the last $\sim
50M$ before merger, the mass increases slightly (seen as a vertical
feature at the right edge of the plot), 
an effect we will discuss in more detail in the context of Fig.~\ref{fig:AntiAlignedInspiralMass}.
The spin is likewise almost constant for $10\lesssim t/M\lesssim 1000$, 
although some 
noise is visible for $t\lesssim 100M$. 

We shall take the steady-state masses and spins evaluated at $t\sim
200M$ as the physical parameters of the binary being studied.
Specifically, all dimensionful quantities will henceforth be expressed
in terms of the mass scale $M$, which we define as the total mass {\em after}
relaxation.

The relaxation of the black holes in the first $\sim 10M$ of the
evolution is also accompanied by the emission of a pulse of unphysical
``junk radiation.''  This pulse passes through the computational
domain, and leaves through the outer boundary after one light crossing
time.  The junk radiation contains short wavelength features, which
are not resolved in the wave zone.  It turns out that the {\em
  constraint damping parameters} $\gamma_0$ and $\gamma_2$ (see
\cite{Lindblom2006}) influence how the unresolved junk radiation
interacts with the numerical grid.  Large constraint damping
parameters enhance the conversion of the outgoing junk radiation (at the
truncation error level) into incoming modes.  This incoming radiation
then lingers for several light-crossing times within the computational
domain, imprinting noise into the extracted gravitational radiation.
For small constraint damping parameters, this conversion is greatly
suppressed, and numerical noise due to junk radiation diminishes much
more rapidly.  The simulations presented here use
$\gamma_0=\gamma_2\sim0.00225/M$ in the wave zone;
these values are smaller by a factor 100 than those used
in~\cite{Boyle2007,Scheel2008}.  (Even smaller constraint damping
parameters fail to suppress constraint violations. Note that
constraint damping parameters are much larger,
$\gamma_0=\gamma_2\sim 3.56/M$, in the vicinity of the black
holes.)  The waveforms presented here show consequently reduced
contaminations in the early part of the evolution
will be discussed in Sec.~\ref{sec:Discussion}, 
cf. Fig.~\ref{fig:0093c_vs_DD}.

\subsection{Inspiral}
\label{sec:Inspiral}

During the inspiral, the mapping between the comoving and inertial frames
is chosen in the same way
as in Refs.~\cite{Boyle2007,Scheel2008} 
and is denoted by ${\cal M}_{\rm I}: x'^i\to x^i$,
where primed coordinates 
denote the comoving frame and unprimed coordinates denote the inertial 
frame. Explicitly, this map is
\begin{eqnarray}
\label{eq:CubicScaleMap}
r      &=& \left[a(t) + \left(1-a(t)\right) \frac{r'^2}{R_0'^2} \right] r', \\
\theta &=& \theta',\\
\phi   &=& \phi' + b(t),\label{eq:CubicScaleMapPhi}
\end{eqnarray}
where $(r,\theta,\phi)$ and $(r',\theta',\phi')$ denote spherical
polar coordinates relative to the center of mass of the system in
inertial and comoving coordinates, respectively.
We choose $R_0'=467M$.
The functions $a(t)$ and $b(t)$ are determined by a dynamical control
system as described in Ref.~\cite{Scheel2006}.
This control system
adjusts $a(t)$ and $b(t)$ so that the centers of the
apparent horizons remain stationary in the comoving frame.  

While each hole is roughly in equilibrium during inspiral, we choose
the gauge source function $H_a$ in the same way as in
Refs.~\cite{Boyle2007,Scheel2008}: A new quantity $\tilde{H}_a$ is
defined that has the following two properties: 1) $\tilde{H}_a$
transforms like a tensor, and 2) in inertial coordinates $\tilde{H}_a
= H_a$. $H_a$ is chosen so that the constraint Eq.~(\ref{e:ghconstr})
is satisfied initially, and $\tilde{H}_{a'}$ is kept constant in the
comoving frame, {\it i.e.,}
\begin{equation}
\partial_{t'} \tilde{H}_{a'} = 0.
\end{equation}
Here primes refer to comoving frame coordinates.
This is essentially an equilibrium condition.

\subsection{Plunge}
\label{sec:Plunge}
We make two key modifications to our algorithm to allow
evolution through merger.
The first is a change in gauge conditions, as in 
Ref.~\cite{Scheel2008}. The second is a change in coordinate mappings 
that allows the excision boundaries to more closely track the
horizons. We describe both of these changes here.

Following Ref.~\cite{Scheel2008}, at some time $t=t_g$ 
(where $g$ stands for ``gauge'') we promote
the gauge source function $H_a$ to an independent dynamical field that
satisfies
\begin{equation}
\label{eq:Hevolution}
\nabla^c\nabla_c H_a = Q_a(x,t,\psi_{ab}) + \xi_2 t^b\partial_b H_a.
\end{equation}
Here $\nabla^c\nabla_c$ is the curved space {\it scalar\/} wave
operator (i.e. each component of $H_a$ is evolved as a scalar),
$\psi_{ab}$ is the spacetime metric, and $t^a$ is the timelike unit normal
to the hypersurface.
The driving functions $Q_a$ are
\begin{eqnarray}
\label{eq:Hevolutiont}
Q_t &=& f(x,t)\xi_1\frac{1-N}{N^\eta},\\
\label{eq:Hevolutioni}
Q_i &=& g(x,t)\xi_3\frac{\beta_i}{N^2},
\end{eqnarray}
where $N$ and $\beta^i$ are the lapse function and the shift vector,
$\eta$, $\xi_1$, $\xi_2$, and $\xi_3$ are constants, and $f(x,t)$, and
$g(x,t)$ are prescribed functions of the spacetime coordinates. 
Eq.~\eqref{eq:Hevolution} is evolved in first-order form, as described in 
Ref.~\cite{Scheel2008}.  Eq.~(\ref{eq:Hevolution}) requires 
values of $H_a$ and
its time derivative as initial data; these are chosen so that
$H_a$ and $\partial_t H_a$ are continuous at $t=t_g$.

This gauge is identical to the one used in Ref.~\cite{Scheel2008},
except that the parameters and functions
that go into Eq.~\eqref{eq:Hevolution} are chosen slightly differently:
We set
$\eta=4$, $\xi_1=0.1$, $\xi_2=6.5$, $\xi_3=0.01$, and 
\begin{eqnarray}
f(x,t) &=& (2-e^{-(t-t_g)/\sigma_1}) \nonumber\\
&\times& (1-e^{-(t-t_g)^2/\sigma_2^2})
e^{-r'^2/\sigma_3^2},\label{eq:GaugeRolloff1}\\
g(x,t) &=& (1-e^{-(t-t_g)/\sigma_4}) \nonumber\\
&\times& (1-e^{-(t-t_g)^2/\sigma_5^2})(t-t_g)
e^{-r'^2/\sigma_3^2},\label{eq:GaugeRolloff2} 
\end{eqnarray}
where $r'$ is the coordinate radius in comoving coordinates, and the
constants are $\sigma_1\sim 62M$, $\sigma_2\sim 44.5M$, 
$\sigma_3\sim 35M$, $\sigma_4\sim 4.5M$, and 
$\sigma_5\sim 3M$. 
The function $g(x,t)$ in $Q_i$, which drives the shift towards 
zero near the black holes, has a factor $(t-t_g)$ that is absent
in Ref.~\cite{Scheel2008}. Prescribing $g(x,t)$ in this way drives the shift
towards zero more strongly at late times, which for this case
is more effective in preventing
gauge singularities from developing.

\begin{figure}
\includegraphics[width=0.45\textwidth]{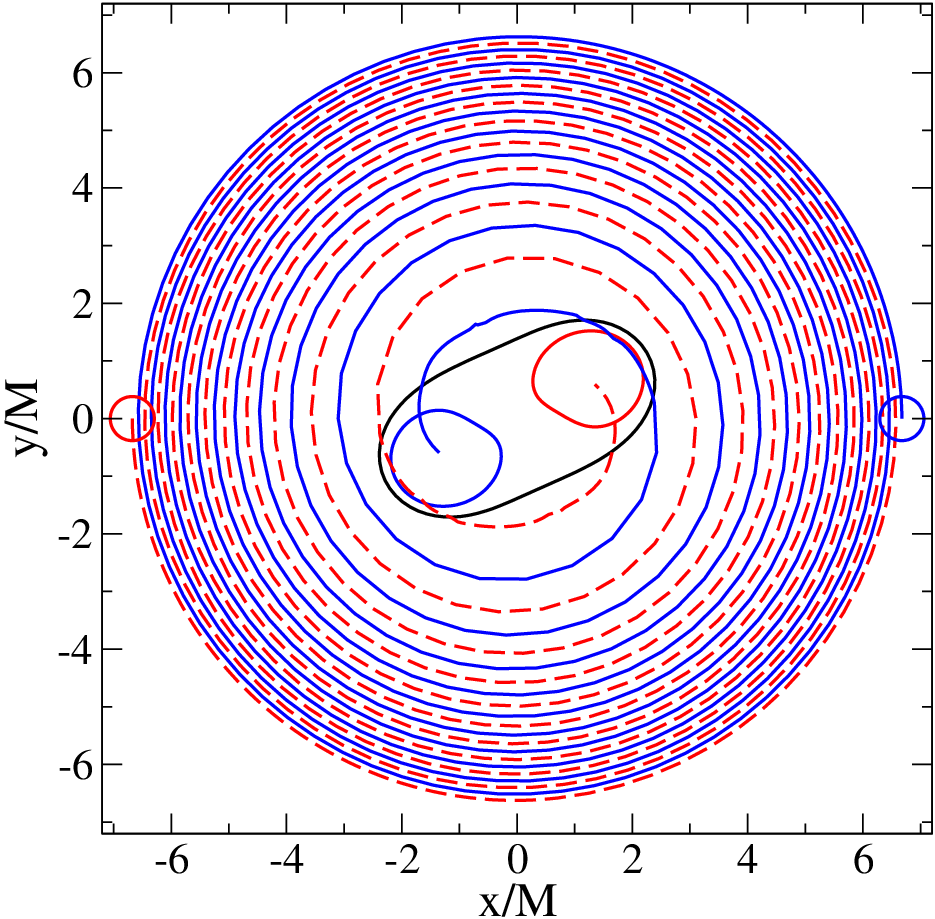}
\caption{\label{fig:AntiAlignedTrajectories}
 Coordinate trajectories of the centers of the apparent horizons 
 represented by the blue and red curves, up until the formation of a common 
 horizon. The closed curves show the coordinate shapes of the corresponding 
 apparent horizons.}
\end{figure}

The second change we make at $t=t_g$ is to control the shape of
each excision boundary so that it matches the shape of the corresponding
apparent horizon.  In the comoving frame, where the excision boundaries
are spherical by construction, this means adjusting the coordinate
mapping between the two frames such that the apparent horizons are also
spherical.  Without this ``shape control'',  the horizons become
sufficiently distorted with respect to the excision boundaries
that the excision boundaries fail
to remain outflow surfaces and our excision algorithm fails.
For the non-spinning black hole binary in Ref.~\cite{Scheel2008},
shape control was not necessary before merger.
To control the shape of black hole 1, we define the map
${\cal M}_{{\rm AH}\,1}: x'^i\to \tilde{x}^i,$ 
\begin{eqnarray}
\tilde{\theta} &=&\theta',\\
\tilde{\phi}   &=&\phi',\\
\tilde{r}      &\equiv& r' - q_1(r')\sum_{\ell=0}^{\ell_{\rm max}}
                          \sum_{m=-\ell}^{\ell} \lambda^1_{\ell m}(t)
                          Y_{\ell m}(\theta',\phi'),
\end{eqnarray}
where
\begin{equation}
q_1(r') = e^{-(r'-r'_0(t))^3/\sigma_q^3},
\end{equation}
and $(r',\theta',\phi')$ are spherical polar coordinates
centered at the (fixed) comoving-coordinate location of black hole
1. The function $q_1(r')$ limits the action of the map to the
vicinity of hole 1.
The constant $\sigma_q$ is chosen to be $\sim 4.5M$, and
$r'_0(t)=r'_0+\nu_1(t-t_g)^{2.1}$ is a function of time that approximately
follows the radius of the black hole, with constants $r'_0 \sim 1.2M$ and 
$\nu_1 \sim 0.00046M$. 
Similarly, we
define the map ${\cal M}_{{\rm AH}\,2}$ for 
black hole 2.
Then the full map ${\cal M}_{\rm m}: x'^i\to x^i$
from the comoving coordinates $x'^i$ to the inertial
coordinates $x^i$ is given by
\begin{equation}
{\cal M}_{\rm m}:=  
     {\cal M}_{\rm I}    \circ
     {\cal M}_{{\rm AH}\,2}  \circ
     {\cal M}_{{\rm AH}\,1}.
\end{equation}
The functions $\lambda^1_{\ell m}(t)$ and $\lambda^2_{\ell m}(t)$ are 
determined by dynamical control systems 
as
described in Refs.~\cite{Scheel2006,Scheel2008}, so that 
the apparent horizons are driven to spheres (up to spherical harmonic 
component $l=l_{\rm max}$) in comoving coordinates.
Note that ${\cal M}_{{\rm AH}\,1}: x'^i\to \tilde{x}^i$ is essentially the same
map that we use to control the shape of the merged horizon during
ringdown, and the control system for that map (and for the map
${\cal M}_{{\rm AH}\,2}$) is the same as the one described in
Ref.~\cite{Scheel2008} for controlling the shape of the merged horizon.

In addition to the modifications to the gauge conditions and coordinate map 
described above, the numerical resolution is also increased slightly 
around the two black holes during this more dynamical phase, and the evolution 
is continued until time $t_m$, 
shortly after the formation of a common horizon.
The coordinate trajectories of the apparent horizon centers 
are shown in Fig.~\ref{fig:AntiAlignedTrajectories} up until $t_m$, 
at which point the binary has gone through 10.6 orbits.

\subsection{Ringdown}
\label{sec:Ringdown}
Our methods for continuing the evolution once a
common horizon has formed are the same as in Ref.~\cite{Scheel2008}.
After a common apparent horizon is found, all variables 
are interpolated onto a new computational domain that has 
only a single excised region. Then, a new comoving coordinate system (and a 
corresponding mapping to inertial coordinates) is chosen so that the new 
excision boundary tracks the shape of the apparent horizon in the inertial 
frame, and also ensures that the outer boundary behaves smoothly in time. 
The gauge conditions are modified as well: the shift vector is no longer 
driven to zero, so that the solution can relax to 
a time-independent state. 
This is done by allowing the gauge function $g(x,t)$ 
that appears in Eq.~(\ref{eq:Hevolutioni}) to gradually approach
zero; the gauge source function $H_a$ still obeys
Eqs.~(\ref{eq:Hevolution}--\ref{eq:Hevolutioni}) as
during the plunge. Specifically,
we change the functions $f(x,t)$ and $g(x,t)$ from
Eqs.~\eqref{eq:GaugeRolloff1} and~\eqref{eq:GaugeRolloff2} to
\begin{eqnarray}
f(x,t) &=& (2-e^{-(t-t_g)/\sigma_1}) \nonumber\\
&\times& (1-e^{-(t-t_g)^2/\sigma_2^2})e^{-r''^2/\sigma_3^2}, \\
g(x,t) &=& (1-e^{-(t-t_g)/\sigma_4}) \nonumber\\
&\times& (1-e^{-(t-t_g)^2/\sigma_5^2})(t-t_g) 
e^{-r''^2/\sigma_3^2} \nonumber\\ 
&\times& e^{-(t-t_m)/\sigma_6^2},
\label{eq:GaugeRolloffRingdown2}
\end{eqnarray}
where $r''$ is the coordinate radius in the new comoving coordinates, 
$\sigma_6\sim 3.1M$, and $t_m$ (here $m$ stands for ``merger'')
is the time we transition to the new domain 
decomposition.

\section{Properties of the numerical solutions}
\label{sec:NumericalProps}

\subsection{Constraints} 
\label{sec:Constraints}
\begin{figure}
\includegraphics[width=0.45\textwidth]{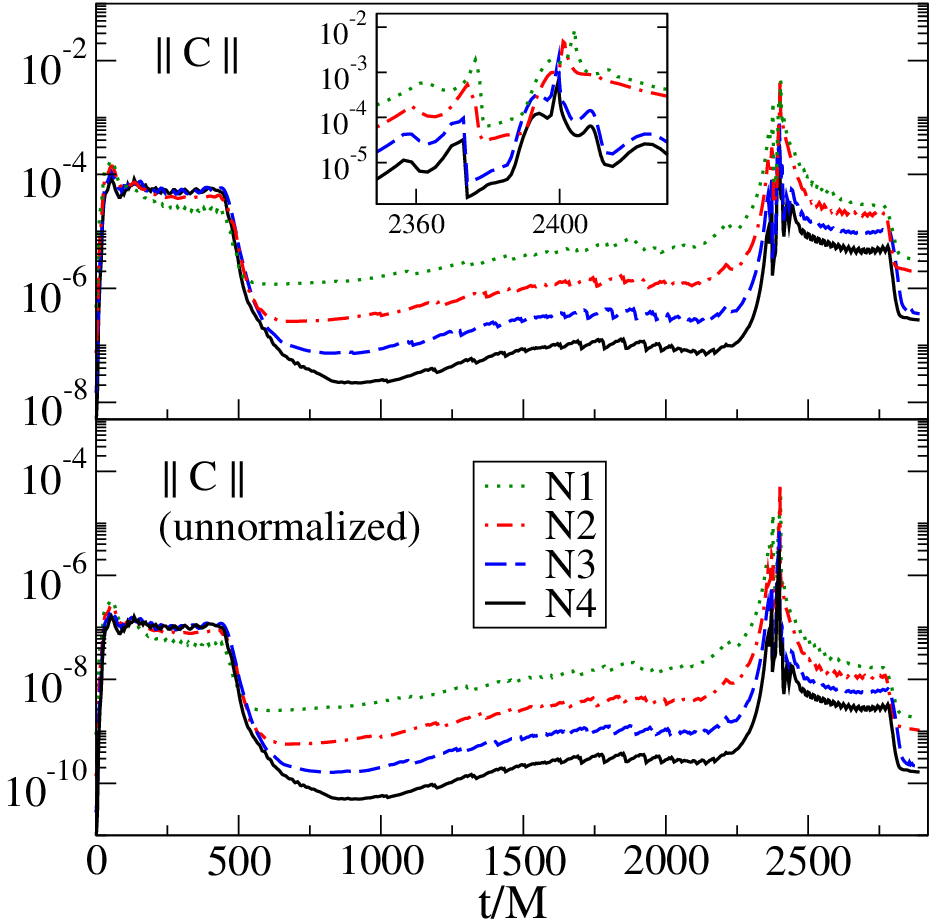}
\caption{\label{fig:AntiAlignedConstraints}
 Constraint violations of runs on different resolutions. The top panel 
 shows the $L^2$ norm of all constraints, normalized by the $L^2$ norm of 
 the spatial gradients of all dynamical fields. The bottom panel shows the 
 same data, but without the normalization factor. The $L^2$ norms are 
 taken over the portion of the computational volume that lies outside 
 apparent horizons.  Note that the time when we change the gauge before
 merger, $t_g\sim 2370M$, and the time when we regrid onto a new single-hole
 domain after merger, $t_m\sim 2400M$, are slightly different for different
 resolutions.}
\end{figure}

We do not explicitly enforce either the Einstein constraints or the secondary 
constraints that arise from writing the system in first-order form. Therefore, 
examining how well these constraints are satisfied provides a useful 
consistency check. Figure~\ref{fig:AntiAlignedConstraints} shows the contraint
violations for the evolutions at different resolutions. The top panel shows 
the $L^2$ norm of all the constraint fields of our first-order generalized 
harmonic system, normalized by the $L^2$ norm of the spatial gradients of 
the dynamical fields (see Eq. (71) of Ref.\cite{Lindblom2006}). The bottom 
panel shows the same quantity, but without the normalization factor 
({\em i.e.}, just the numerator of Eq. (71) of Ref.\cite{Lindblom2006}). 
The $L^2$ norms are taken over the portion of the computational volume that 
lies outside the apparent horizons.

The constraints increase as the black holes approach each other and become 
increasingly distorted. At $t_g=2372.05M$ for N4 ($t_g=2372.05M$ for 
N3, $t_g=2376.5M$ for N2, $t_g=2376.5M$), the gauge conditions are 
changed (cf.~\ref{sec:Plunge}) and the resolution around the holes is 
increased slightly. Because of the change in resolution, the constraints drop 
by more than an order of magnitude. Close to merger, the constraints grow 
larger again. The transition to a single-hole evolution 
(cf.~\ref{sec:Ringdown}) occurs at $t_m=2399.64M$ for N4 ($t_m=2399.66M$ 
for N3, $t_m=2401.27M$ for N2, $t_m=2404.23M$ for N1). 
Shortly after this time, the constraints drop by about 
two orders of magnitude. 
This is because the largest constraint violations occur near and between
the individual apparent horizons, and this region is newly excised from
the computational domain at $t=t_m$.

\subsection{Black hole spins and masses} 
\label{sec:BHSpinsMasses}

\begin{figure}
\includegraphics[width=0.45\textwidth]{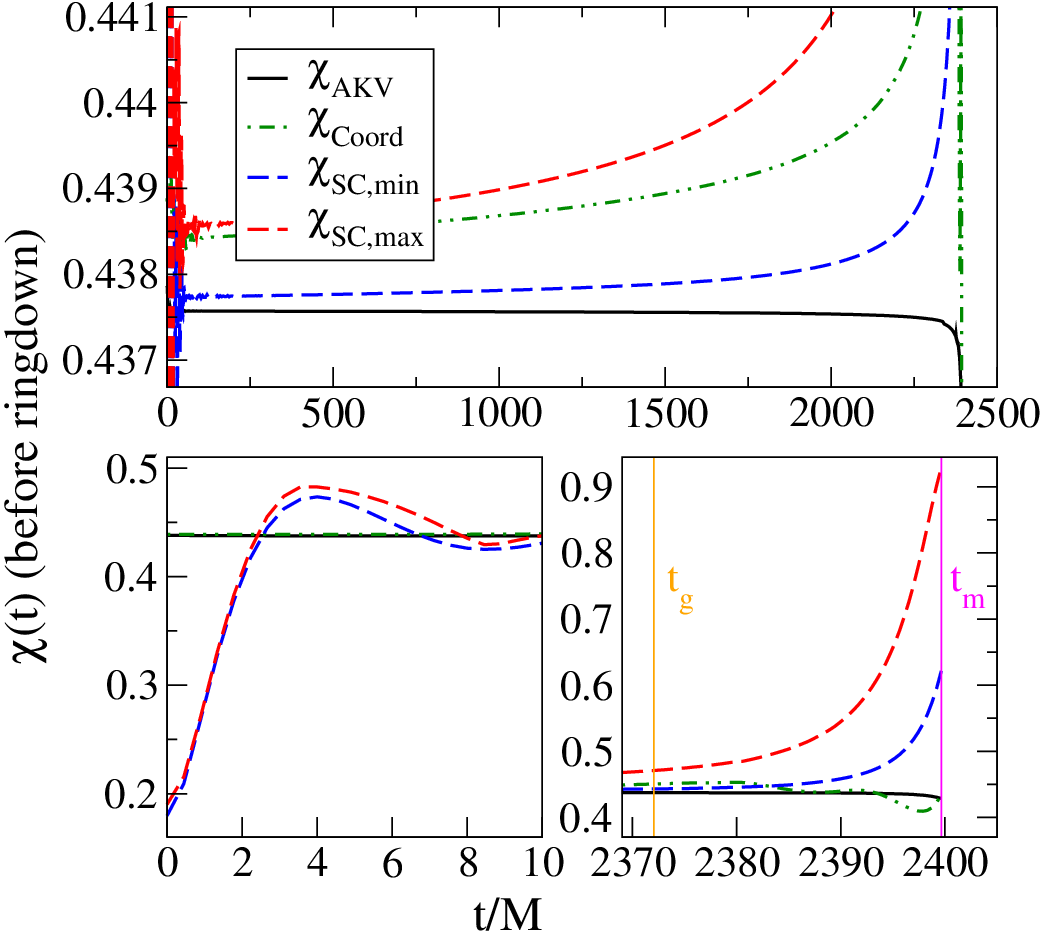}
\caption{\label{fig:AntiAlignedInspiralSpins}
 Dimensionless spins $\chi$ of one black hole in the N4 evolution,
 evaluated using an approximate Killing vector, a coordinate rotation
 vector $-\partial_\phi$, or the extrema of the instrinsic scalar curvature
 on the apparent horizon.
Bottom panels show detail at early and late times. Also shown
are the time of gauge change $t_g$ before merger, and the time $t_m$ that
we transition to a single-hole evolution just after merger.}
\end{figure}

There are different ways to compute the spin $\chi(t)$ of a black hole. 
The approach we prefer computes the
spin from an angular momentum surface 
integral~\cite{BrownYork1993,Ashtekar-Krishnan:2004}
using approximate Killing vectors of the apparent horizons, as
described in~\cite{OwenThesis,Lovelace2008} (see
also~\cite{Dreyer2003,Cook2007}).
We shall denote the resulting spin 
by $\SMM_{\text{AKV}} (t)$. Another less sophisticated method simply uses 
coordinate rotation vectors, and we denote the resulting spin by 
$\SMM_{\text{Coord}} (t)$. We also use 
two more spin diagnostics that are based on 
the minimium and maximum of the instrinsic scalar curvature of the 
apparent horizon for a Kerr black hole~\cite{Lovelace2008}; we call these
$\SpinFromShapeMin (t)$ and $\SpinFromShapeMax (t)$. 
These last two measures of spin are
expected to give reasonable results when the black holes 
are sufficiently far apart and close to equilibrium, and
after the final black hole has settled down to a time-independent state.
However, they are expected to be less accurate near 
merger and at the start of the evolution.

\begin{figure}
\includegraphics[width=0.45\textwidth]{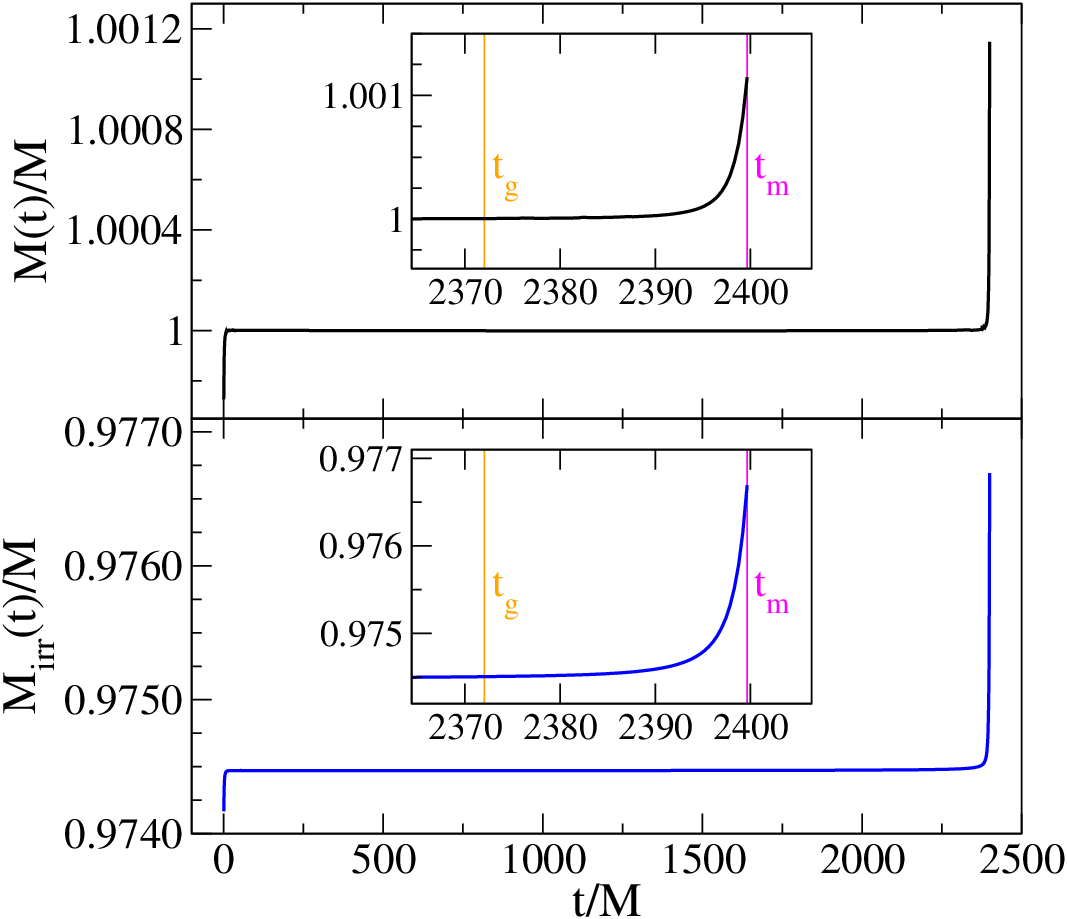}
\caption{\label{fig:AntiAlignedInspiralMass} 
Sum of Christodoulou masses $M(t)$ and sum of
irreducible masses $M_{\rm irr}(t)$ of the two black holes
during inspiral.  The data is from the N4 evolution, and uses
$\chi_{\textrm{AKV}}$ when computing $M(t)$.
Insets show detail at late times, and indicate the
transition times $t_g$ and $t_m$.  
}
\end{figure}

Fig.~\ref{fig:AntiAlignedInspiralSpins} shows 
these four spin measures for 
black hole 1 in the N4 evolution during inspiral and 
plunge.
From the lower left panel we see that $\SpinFromShapeMin (t)$ and
$\SpinFromShapeMax (t)$ differ from $\chi_{\rm Coord} (t)$ and 
$\chi_{\rm AKV} (t)$ by more than a factor of two at $t=0$.  
This indicates that the initial black holes
do not have the appropriate shape
for the Kerr solution; i.e. they are
distorted because of the way the initial data is constructed.
As the black holes relax, $\SpinFromShapeMin (t)$
and $\SpinFromShapeMax (t)$ approach the other two spin measures.  
The relaxed spin at $t\sim200M$ is $\chi=0.43757 \pm 0.00001$, 
where the uncertainty is based on the variation in $\chi_{\rm AKV}$ 
between $t=100M$ and $t=1000M$.
During the inspiral, $\chi_{\rm AKV}(t)$ decreases slowly and 
monotonically, dropping by $10^{-4}$
at $90M$ before merger, and dropping by $0.01$ at the time of merger.
Tidal dissipation should {\em slow down} the black holes, so this
decrease is physically sensible.  In contrast, the other three spin
diagnostics show a mild {\em increase} in spin, 
suggesting that they are
less reliable.
Close to merger, $\SpinFromShapeMin (t)$ and 
$\SpinFromShapeMax (t)$ increase dramatically, with $\SpinFromShapeMax (t)$ 
growing as large as 0.92.  In this regime, the shapes of the
individual black holes are dominated by tidal distortion, and are therefore
useless for measuring the spin.

The Christodoulou mass $m$ of one black hole, as defined in 
Eq.~(\ref{eq:Christodoulou}), depends on the spin. 
We take $\chi_{\text{AKV}}(t)$ as the preferred spin measure, and use it to 
compute the total Christodoulou mass $M(t)$ 
during the inspiral and plunge. This is shown in the top panel of 
Fig.~\ref{fig:AntiAlignedInspiralMass}. 
The Christodoulou mass settles 
down to $M(t)/M=1.000000$ after $t=150M$ (this defines $M$), 
and increases to 
$M(t)/M=1.00114$ at the time of merger.  Most 
of the increase in mass occurs very close to merger, 
as can be seen from the inset of Fig.~\ref{fig:AntiAlignedInspiralMass}.  
Until about $30M$ before merger 
(i.e. $t=2370M$), the mass is constant to a few parts in $10^6$. 
For comparison, in the bottom panel we also 
display $M_{\text{irr}}(t)$, the sum of the irreducible masses, which does 
not depend on the spin. This quantity 
settles down to 
$M_{\text{irr}}(t)/M=0.974508$ at $t=200M$, and increases to 
$M_{\text{irr}}(t)/M=0.97668$ at $t=2400M$.  
Again, almost all of this increase happens shortly before merger.
During the inspiral up to $30M$ before merger, $M_{\text{irr}}(t)/M$
increases by only $6\times 10^{-5}$, but in the last $30M$ the
increase is $\sim 0.002$.

\begin{figure}
\includegraphics[width=0.45\textwidth]{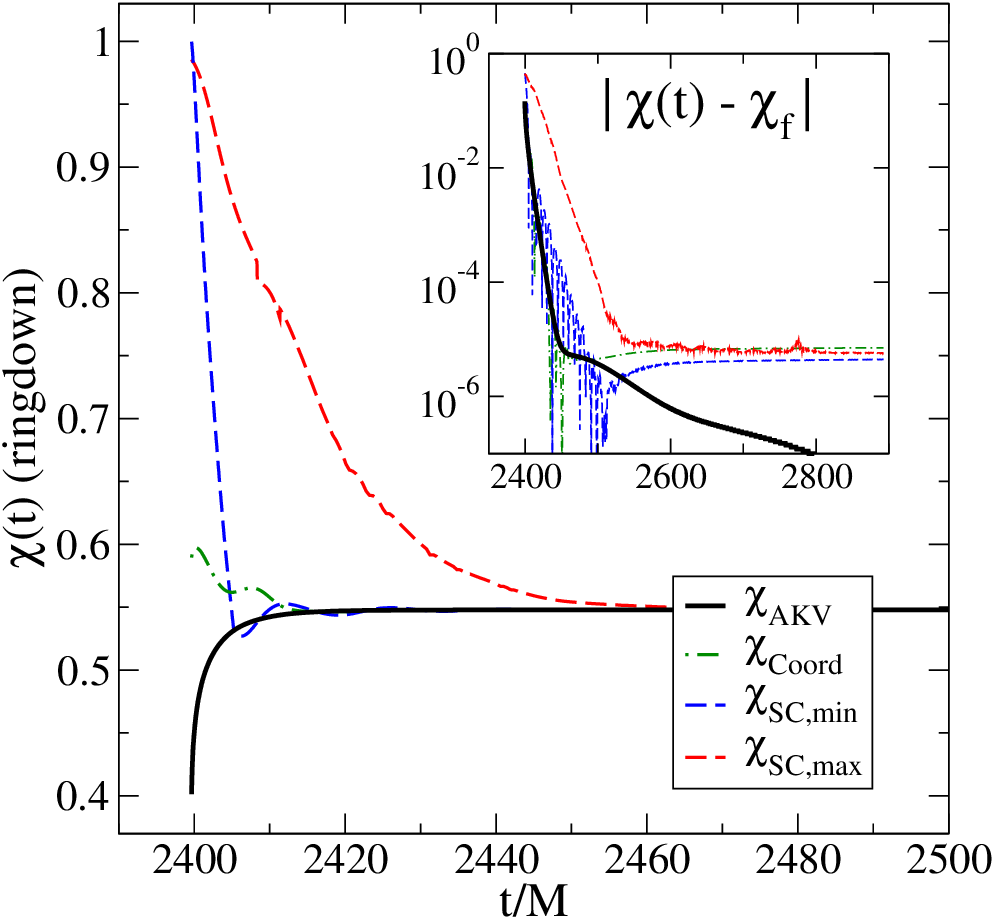}
\caption{\label{fig:AntiAlignedSpinsRingdown} Dimensionless
    spins $\chi(t)$ of the final black hole in the N4 evolution.  The
    (most reliable) spin diagnostic $\chi_{\rm AKV}$ starts at $\sim
    0.4$ and increases to its final value $\chi_f=0.54781 \pm 0.00001$. 
    The other spin diagnostics are unreliable for the highly distorted 
    black hole shortly after merger, but subsequently approach
    $\chi_{\rm AKV}$.  
}
\end{figure}

The merger results in one highly distorted black hole, which
subsequently rings down into a stationary Kerr black hole.
Figure~\ref{fig:AntiAlignedSpinsRingdown} shows our four spin
diagnostics during the ringdown.  The spin measures
$\SpinFromShapeMin (t)$ and $\SpinFromShapeMax (t)$ assume a Kerr black hole.
Just after merger, the horizon is highly distorted, so these two spin
diagnostics are not valid there.  However, as the remnant black hole
rings down to Kerr, $\SpinFromShapeMax (t)$ and $\SpinFromShapeMin (t)$
approach the quasi-local AKV spin to better than 1 part in $10^5$ (see
the inset of Fig.~\ref{fig:AntiAlignedSpinsRingdown}).  The
quasi-local spin based on coordinate rotation vectors, $\chi_{\rm
  Coord} (t)$, also agrees with the other spin measures to a similar level
at late times. The spin of the final black hole points in the 
direction of the initial orbital angular momentum.

\begin{figure}
\includegraphics[width=0.46\textwidth]{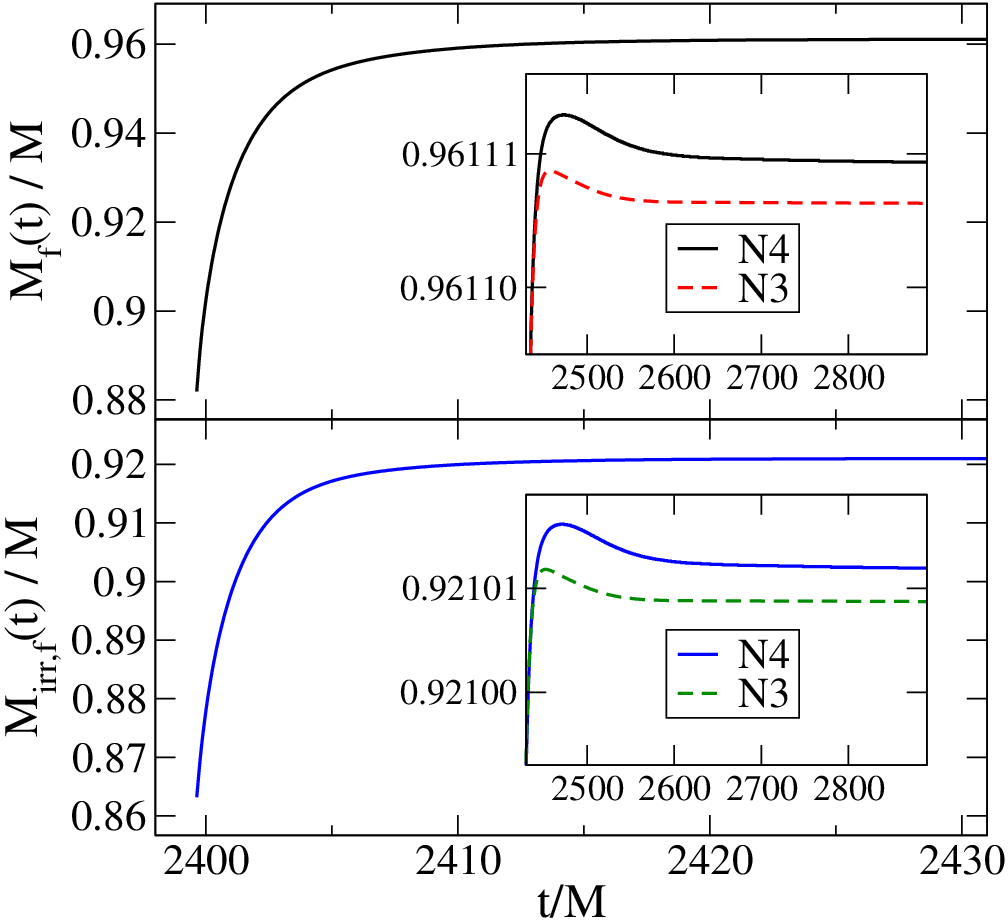}
\caption{\label{fig:AntiAlignedRingdownMass}
 The top panel shows the Christodoulou mass $M_f(t)$ of the final black hole 
 in the N4 and N3 runs, computed using $\chi_{\text{AKV}}(t)$. The bottom 
 panel shows the irreducible mass $M_{\text{irr,f}}(t)$. 
}
\end{figure}

The Christodoulou mass $M_f(t)$ of the final 
black hole in the N4 evolution, again evaluated using $\chi_{\text{AKV}}(t)$, 
is shown in the top panel of Fig.~\ref{fig:AntiAlignedRingdownMass}. 
The mass settles down to a final value of 
$M_f/M=0.961109 \pm 0.000003$. The bottom panel shows the 
irreducible mass $M_{\text{irr,f}}(t)$ of the final black hole, which settles 
down to a final value of $M_{\text{irr,f}}=0.921012 \pm 0.000003$. 
The uncertainties are determined from the difference between runs N4 and 
N3, so they include only numerical truncation
error and not any systematic effects.  The uncertainty in the mass is
visible in the insets of Fig.~\ref{fig:AntiAlignedRingdownMass}.

\section{COMPUTATION OF THE WAVEFORM}
\label{sec:Waveform}

\subsection{Waveform extraction}
\label{sec:WaveformExtraction}
Gravitational waves are extracted from the simulution on spheres of 
different values of the coordinate radius $r$, following the same 
procedure as in 
Refs.~\cite{Pfeiffer-Brown-etal:2007,Boyle2008,Scheel2008}.
The Newman-Penrose scalar $\Psi_4$ 
in terms of spin-weighted spherical harmonics of weight -2:
\begin{equation}
\Psi_4\left(t,r,\theta,\phi\right)=\displaystyle\sum_{lm} 
\Psi_4^{lm}\left(t,r\right)_{-2}Y_{lm}\left(\theta,\phi\right),
\end{equation}
where the $\Psi_4^{lm}$ are expansion coefficients defined by this equation.
Here we also focus on the dominant $(l,m)=(2,2)$ mode, and split the 
extracted waveform into real phase $\phi$ and real amplitude $A$,
defined by (see e.g.~\cite{Baker2006a,Bruegmann2006})
\begin{equation}
\label{eq:GWPhaseAmp}
\Psi_4^{22}(r,t)=A(r,t)e^{-i\phi(r,t)}.
\end{equation}
The gravitational-wave frequency is given by
\begin{equation}
\omega = \frac {d\phi}{dt}.
\end{equation}
The minus sign in Eq.~\eqref{eq:GWPhaseAmp} is chosen so that the phase
increases in time and $\omega$ is positive.

\begin{figure}
\includegraphics[scale=0.5]{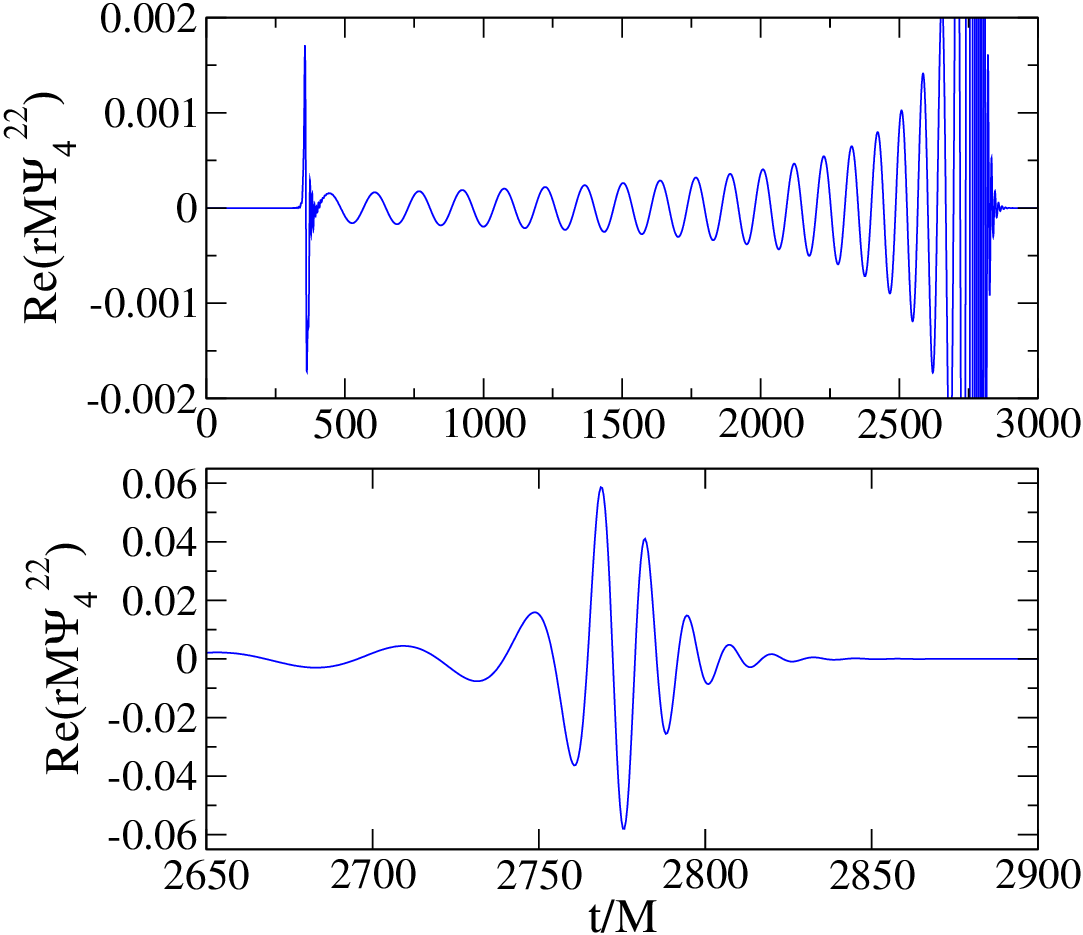}
\caption{\label{fig:Waveform_22_R0350m_Lev6}
 Gravitational waveform extracted at finite radius $r=350M$ 
 for the N4 evolution. The top panel zooms in on the inspiral waveform, 
 and the bottom panel zooms in on the merger and ringdown.}
\end{figure}

The coordinate radius of our outer boundary is located at
$R_{\rm max}=427M$ at $t=0$ and $R_{\rm max}=365M$ at 
$t>2500M$; it shrinks slightly during the evolution because of the
mappings [cf. Eq.~(\ref{eq:CubicScaleMap})] used in our dual frame approach.
The $(l,m)=(2,2)$ waveform, extracted at a single coordinate 
radius $r=350M$ for the N4 evolution, is shown in 
Fig~\ref{fig:Waveform_22_R0350m_Lev6}. The short pulse at $t\sim 360M$ is 
due to junk radiation.  The magnitude of this pulse is about twice 
as large as for non-spinning black holes, cf. Ref.~\cite{Boyle2007,Scheel2008}.

\subsection{Convergence of extracted waveforms}
\label{sec:WaveformConvergence}
In this section we examine the convergence of the gravitational waveforms 
extracted at fixed radius, without extrapolation to infinity. This allows us 
to study the behavior of our code without the complications of extrapolation. 
The extrapolation process and the resulting extrapolated waveforms are 
discussed in Sec.~\ref{sec:Extrapolation}.

\begin{figure}
\includegraphics[scale=0.49]{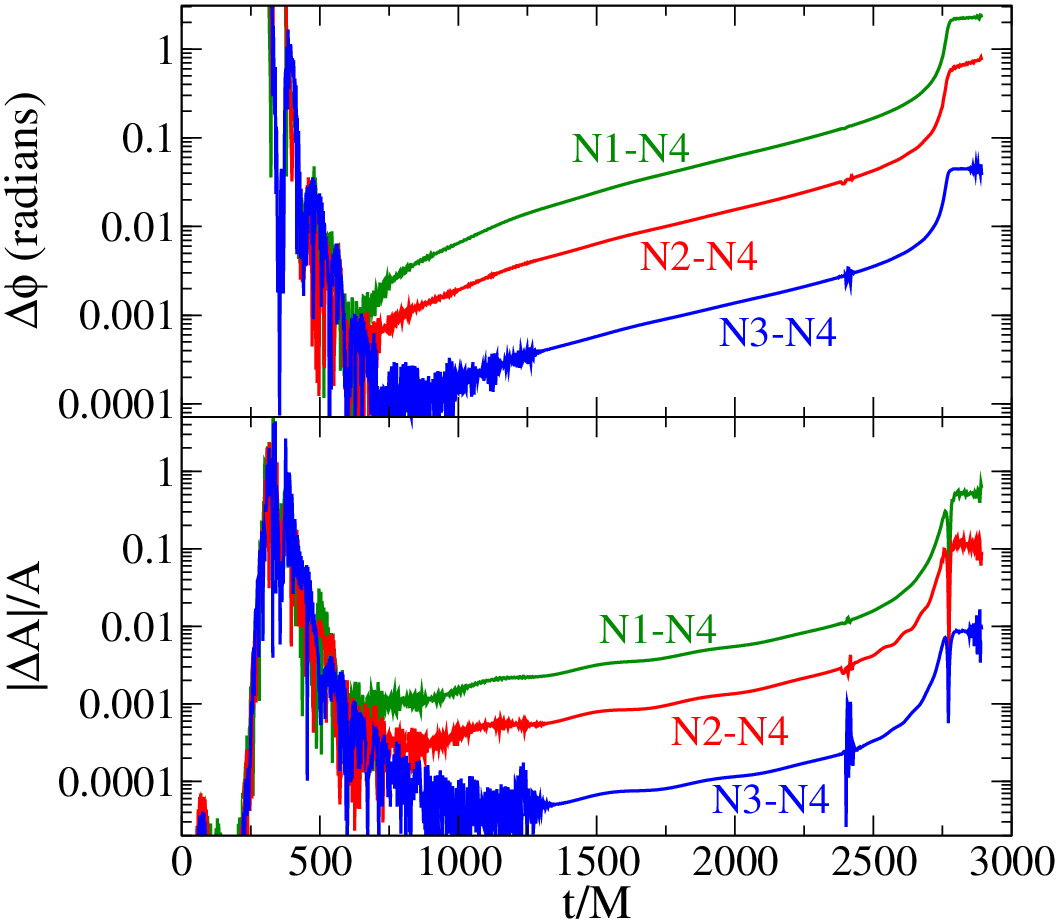}
\caption{\label{fig:AmpPhaseConvergence_22_R0350m}
 Convergence of gravitational waveforms with numerical resolution. Shown 
 are phase and amplitude differences between numerical waveforms $\Psi_4^{22}$ 
 computed using different numerical resolutions. All waveforms are extracted 
 at $r=350M$, and no time shifting or phase shifting is done to align 
 waveforms.}
\end{figure}

Figure~\ref{fig:AmpPhaseConvergence_22_R0350m} shows the convergence of the 
gravitational-wave phase $\phi$ and amplitude $A$ with numerical resolution.
For this plot, the waveform was extracted at a fixed inertial-coordinate 
radius of $r=350M$. Each line in the top panel shows the absolute difference 
between $\phi$ computed at some particular resolution and $\phi$ computed from 
our highest resolution N4 run. The curves in the bottom panel similarly show 
the {\em relative} amplitude differences. When subtracting results at 
different resolutions, no time or phase adjustment has been performed. The 
noise at early times is due to junk radiation generated near $t=0$. Most 
of this junk radiation leaves through the outer boundary after one crossing 
time. The plots show that the phase difference accumulated over 10.6 orbits 
plus merger and ringdown---in total 31 gravitational wave cycles---is 
less than 0.1 radians, and the 
relative amplitude differences are less than 0.017. These
numbers can be 
taken as an estimate of the numerical truncation error of our N3 run.  
Because of the rapid convergence of the code, we expect that the
errors of the N4 run are significantly smaller.

\begin{figure}
\includegraphics[scale=0.49]{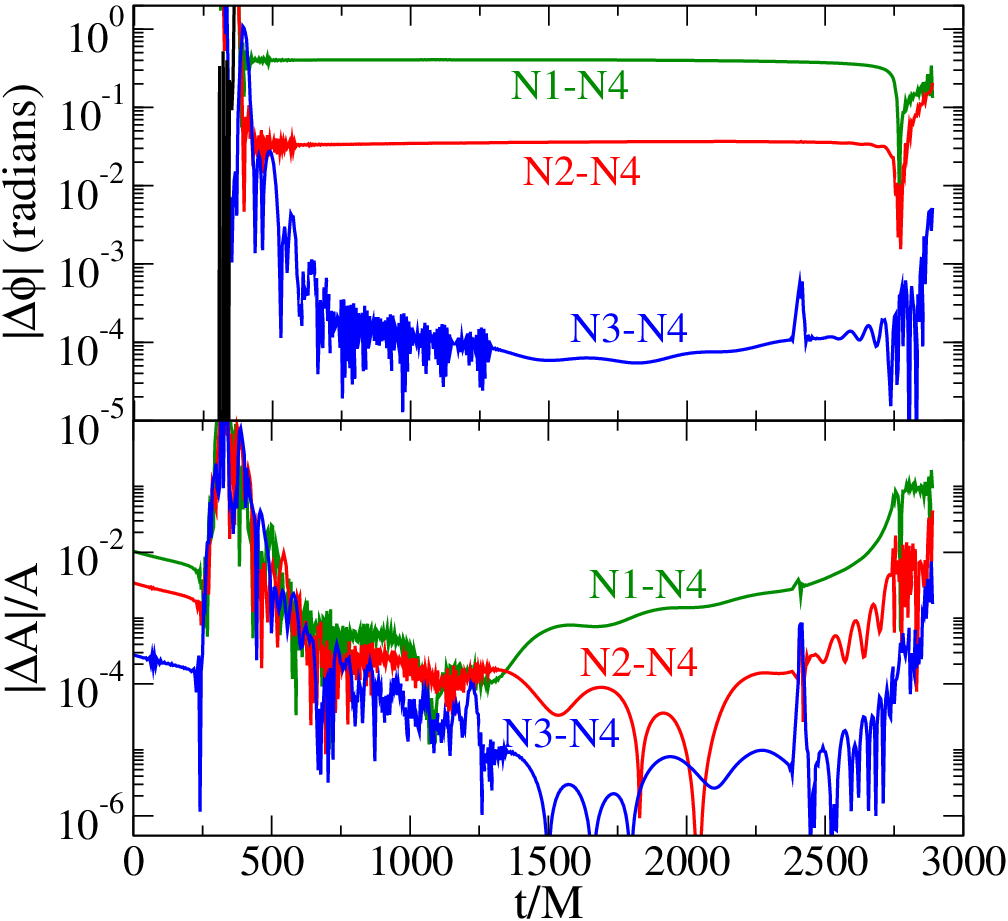}
\caption{\label{fig:AmpPhaseConvergence_22_R0350m_TimePhaseShifted}
 Convergence of gravitational waveforms with numerical resolution. 
 Same as Fig.~\ref{fig:AmpPhaseConvergence_22_R0350m} except all other 
 waveforms are time-shifted and phase-shifted to best match the waveform of 
 the N4 run.}
\end{figure}

Figure~\ref{fig:AmpPhaseConvergence_22_R0350m_TimePhaseShifted} 
is the same as Fig.~\ref{fig:AmpPhaseConvergence_22_R0350m} 
after the N1, N2, N3 
waveforms have been time shifted and phase shifted to best match the waveform 
of the N4 evolution. This best match is determined by a simple least-squares 
procedure: we minimize the function
\begin{equation}
\label{eq:LeastSquaresFit}
\sum_i \left(A_1(t_i)e^{i\phi_1(t_i)}%
-A_2(t_i+t_0)e^{i(\phi_2(t_i+t_0)+\phi_0)}\right)^2,
\end{equation}
by varying $t_0$ and $\phi_0$. Here $A_1$, $\phi_1$, $A_2$, and $\phi_2$ are 
the amplitudes and phases of the two waveforms being matched, and the sum 
goes over all times $t_i$ at which waveform 1 is sampled. This type of 
comparison is relevant for analysis of data from gravitational-wave detectors: 
when comparing experimental data with numerical detection templates, the 
template will be shifted in both time and phase to best match the data. For 
this type of comparison, 
Fig.~\ref{fig:AmpPhaseConvergence_22_R0350m_TimePhaseShifted} shows that the 
numerical truncation error of our N3 run is less than 0.01 radians in phase 
and 0.1 percent in amplitude for $t>550M$. At earlier times, the errors are 
somewhat larger and are dominated by residual junk radiation.

\subsection{Extrapolation of waveforms to infinity}
\label{sec:Extrapolation}
Gravitational-wave detectors measure waveforms as seen by an observer 
effectively infinitely far from the source. Since our numerical 
simulations cover only a finite spacetime volume, after extracting 
waveforms at multiple finite radii, we extrapolate these waveforms
to infinite radius using the procedure described in~\cite{Scheel2008} 
(see also~\cite{Boyle2008} for more details). 
This is intended to reduce near-field effects as well as gauge effects 
that can be caused by the time dependence of the lapse function or 
the nonoptimal choice of tetrad for computing $\Psi_4$.

The extrapolation of the extracted waveforms involves first computing 
each extracted waveform as a function of retarded time $u=t_s-r^*$ and 
extraction radius $r_{\text{areal}}$ (see ~\cite{Scheel2008} for precise 
definitions). Then at each value of $u$, the phase and amplitude are fitted 
to polynomials in $1/r_{\text{areal}}$:
\begin{eqnarray}
\label{eq:ExtrapolatedPhase}
\phi(u,r_{\text{areal}}) &=&
\phi_{(0)}(u)+\sum_{k=1}^n \frac {\phi_{(k)}(u)}{r_{\text{areal}}^k}, \\
\label{eq:ExtrapolatedAmp}
r_{\text{areal}} &=& A_{(0)}(u) + \sum_{k=1}^n \frac {A_{(k)}(u)}{r_{\text{areal}}^k}.
\end{eqnarray}
The phase and amplitude of the desired asymptotic waveform are thus given by
the leading-order term of the corresponding polynomial, as a function of
retarded time:
\begin{eqnarray}
\phi(u) &=& \phi_{(0)}(u), \\
r_{\text{areal}} A(u) &=& A_{(0)}(u).
\end{eqnarray}

\begin{figure}
\centerline{\includegraphics[width=0.49\textwidth]{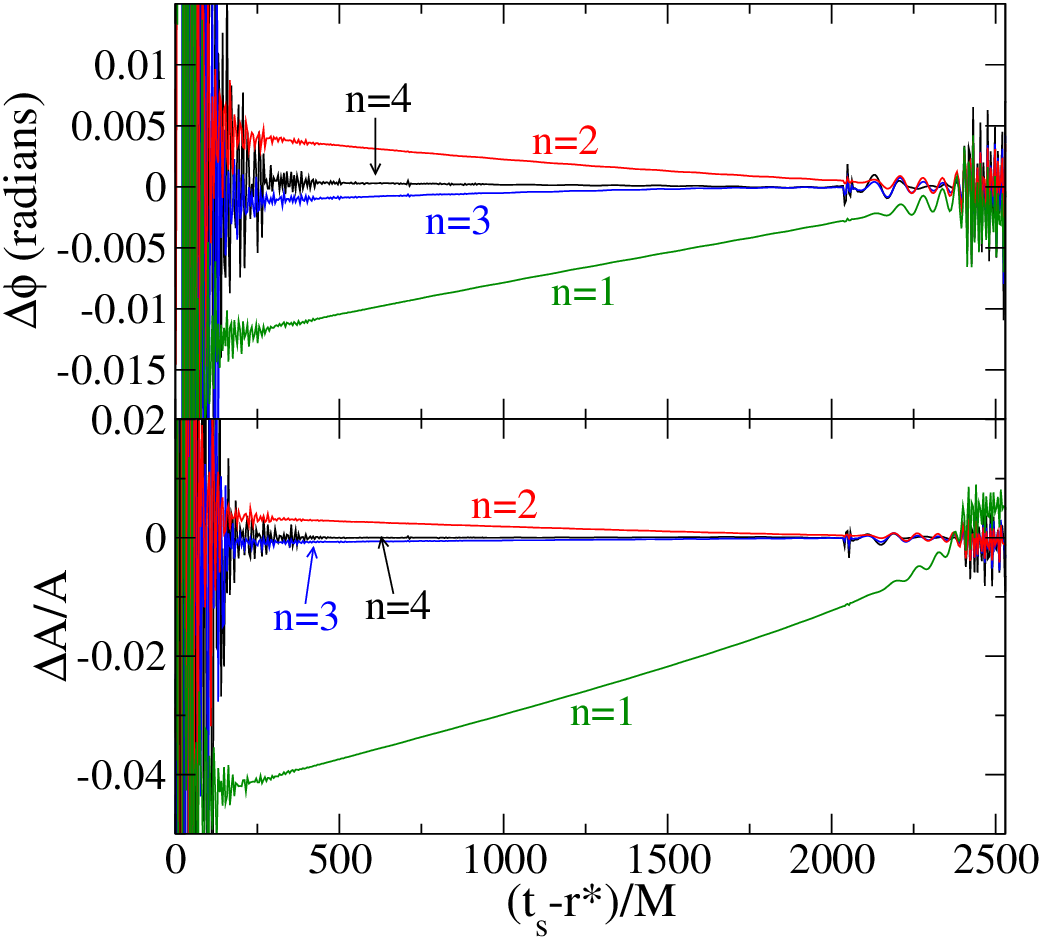}}
\caption{\label{fig:AmpPhaseConvergence_22_Extrapolated}
 Convergence of extrapolation to infinity for extrapolation of order $n$.
 For each $n$, plotted is the extrapolated waveform from N4 using order 
 $n+1$ minus the extrapolated waveform using order $n$. The top panel shows 
 phase differences, the bottom panel shows amplitude differences. No 
 shifting in time or phase has been done for this comparison.}
\end{figure}

Figure~\ref{fig:AmpPhaseConvergence_22_Extrapolated} shows phase and 
amplitude differences between extrapolated waveforms that are computed 
using different values of polynomial order $n$ in 
Eqs.~\eqref{eq:ExtrapolatedPhase} and~\eqref{eq:ExtrapolatedAmp}.
The extrapolation is based on waveforms extracted at 20 different radii
between $75M$ and $350M$.  As in~\cite{Scheel2008}, our preferred 
extrapolation order is
 $n=3$, which gives a phase error of less than 0.004 radians 
and a relative amplitude error of less than 0.006 during most of the 
inspiral, and a phase error of less than 0.01
radians and a relative amplitude error of 0.006 in the ringdown.

\begin{figure}
\includegraphics[width=0.46\textwidth]{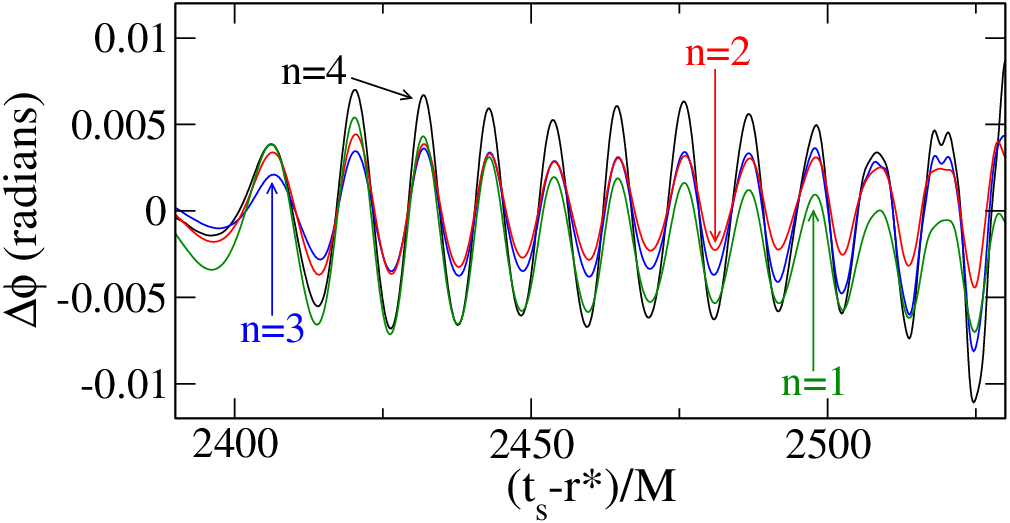}
\caption{\label{fig:PhaseConvergence_22_Extrapolated}
 Late-time phase convergence of extrapolation to infinity. Same as the top 
 panel of Fig.~\ref{fig:AmpPhaseConvergence_22_Extrapolated}, except 
 zoomed to late times. The peak amplitude of the waveform occurs at 
 $t_s-r^*=2410.6M$.}
\end{figure}

\begin{figure}
\includegraphics[width=0.46\textwidth]{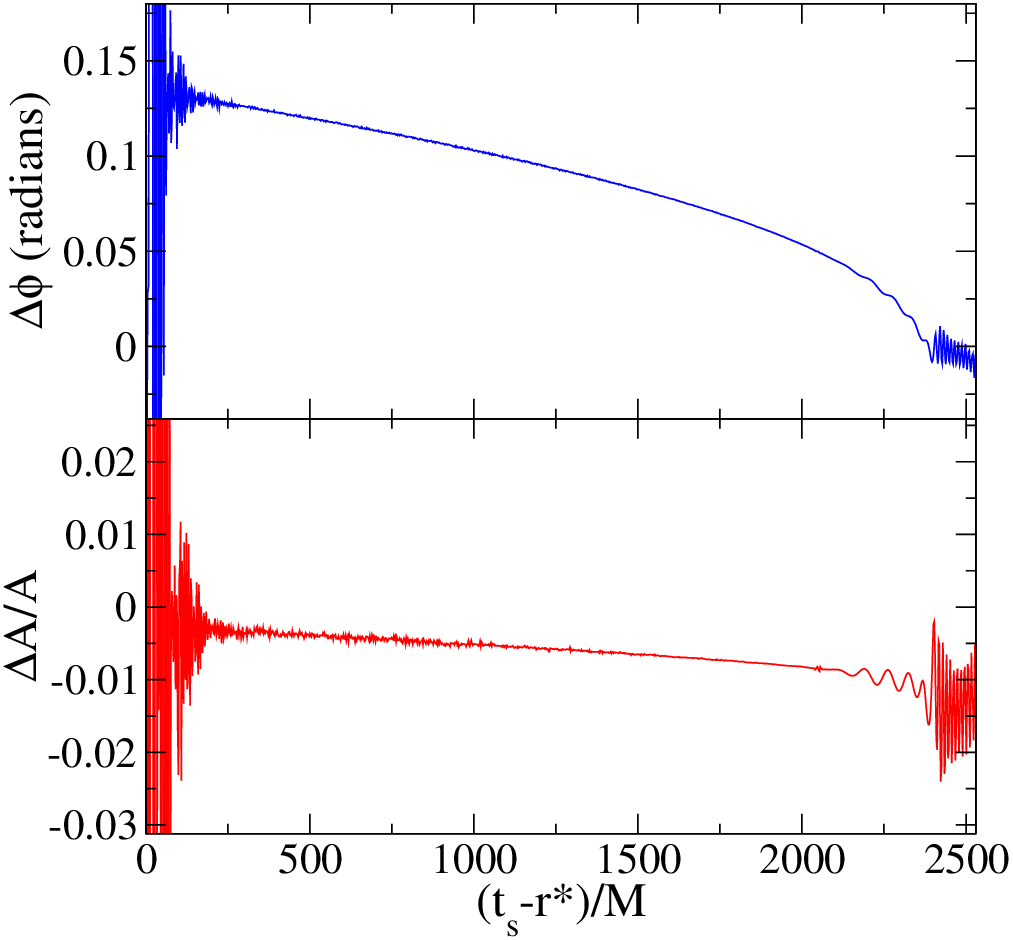}
\caption{\label{fig:AmpPhaseConvergence_ExtrapolatedExtracted}
 Phase and relative amplitude differences between extrapolated and 
 extracted waveforms for N4. The extracted waveform is extracted at 
 coordinate radius $r=350M$. The waveforms are time-shifted and phase-shifted 
 to produce the best least-squares match.}
\end{figure}

Figure~\ref{fig:PhaseConvergence_22_Extrapolated} is the same as the top panel 
of Fig.~\ref{fig:AmpPhaseConvergence_22_Extrapolated}, except zoomed to late 
times. During merger and ringdown, the extrapolation procedure does not 
converge with increasing extrapolation order $n$: the phase differences are 
slightly larger for larger $n$. This was also seen for the extrapolated 
waveforms of our equal-mass nonspinning black hole binary~\cite{Scheel2008}, 
and is possibly due to gauge effects that do not obey the fitting formulae, 
Eqs.~\eqref{eq:ExtrapolatedPhase} and~\eqref{eq:ExtrapolatedAmp}.

\begin{figure*}
\includegraphics[width=0.92\textwidth]{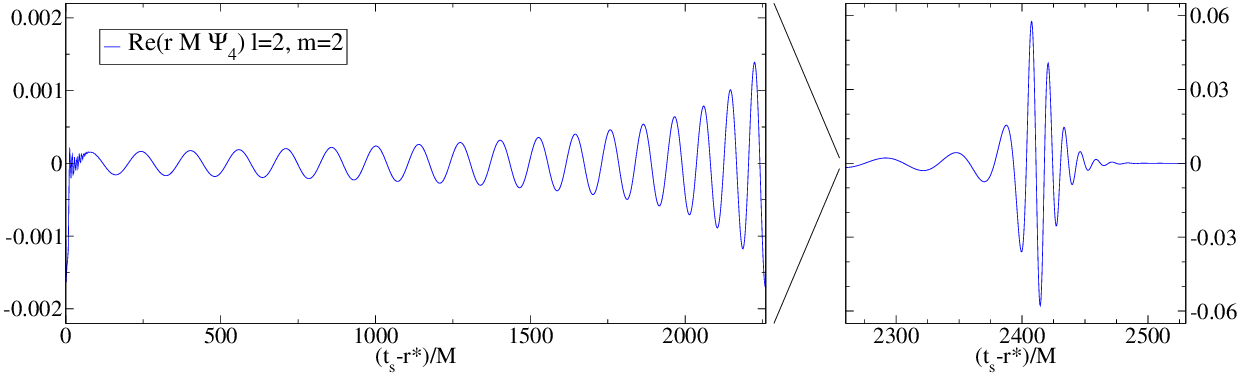}\\[1em]

\includegraphics[width=0.92\textwidth]{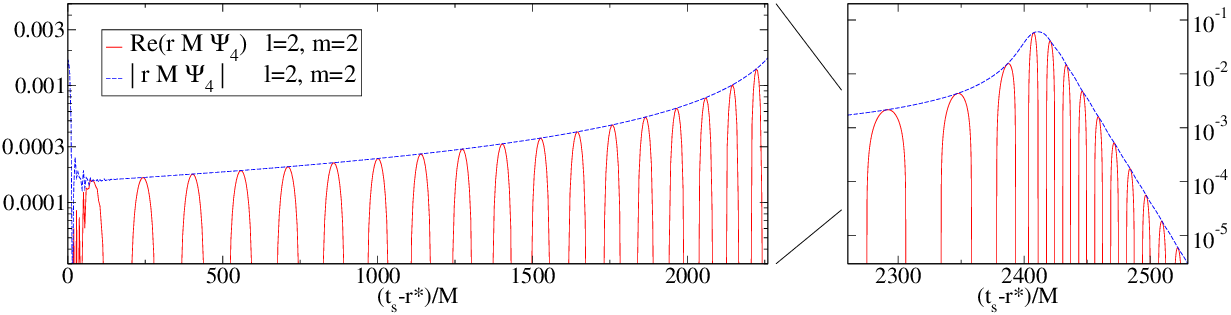}
\caption{\label{fig:ExtrapolatedWaveform}
  Final waveform, extrapolated to infinity.  The top panels
  show the real part of $\Psi_4^{22}$ with a linear y-axis, the bottom
  panels with a logarithmic y-axis.  The right panels show an
  enlargement of merger and ringdown.  }
\end{figure*}

Figure~\ref{fig:AmpPhaseConvergence_ExtrapolatedExtracted} shows the phase 
and amplitude differences between our preferred extrapolated waveform 
using $n=3$ and the waveform extrapolated at coordinate radius $r=350M$, 
both for the N4 run. The extrapolated waveform has been shifted in time 
and phase so as to best match the $n=3$ extrapolated waveform, using the 
least-squares fit of Eq.~\eqref{eq:LeastSquaresFit}. 
The phase difference between extrapolated waveform and waveform
extracted at $r=350M$ becomes as large as $0.13$ radians, and the
amplitude difference
is on the order of 1 per cent.

Figure~\ref{fig:ExtrapolatedWaveform} presents the final waveform after 
extrapolation to infinite radius. There are 22 gravitational-wave cycles 
before the maximum of $|\Psi_4|$, and 9 gravitational-wave cycles during 
ringdown, over which the amplitude of $|\Psi_4|$ drops by four orders of 
magnitude.

\section{Discussion}
\label{sec:Discussion}

We have presented the first spectral computation of a binary black hole
inspiral,
merger, and ringdown with spinning black holes, and find that we can
achieve similar accuracy for the final mass, final spin, and
gravitational waveforms as in the non-spinning case~\cite{Scheel2008}.
For initial spins of $\chi=0.43757\pm 0.00001$, the mass and spin of the 
final hole are $M_f/M=0.961109 \pm 0.000003$ and 
$\chi_f=0.54781 \pm 0.00001$.  The uncertainties are
based on comparing runs at our highest two resolutions, and do not
take into account systematic errors (e.g. the presence of a finite
outer boundary or gauge effects).  Note that for the non-spinning
case~\cite{Scheel2008}, we found that changing the outer boundary
location produced a smaller effect on the final mass and spin than
changing the resolution, and that the outer boundary for the
evolutions presented here is more distant (at late times, when
most of the radiation passes through the boundary) than it was in
Ref.~\cite{Scheel2008}.  The uncertainties in the gravitational
waveforms are $\lesssim 0.01$ radians in phase and 
$\lesssim 0.6$ percent in amplitude (when waveforms are time and phase
shifted). These uncertainties are based on comparisons between our two
highest resolution runs and comparisons between different methods of
extrapolating waveforms to infinite extraction radius.

The methods used here to simulate plunge and ringdown are similar to
those in Ref.~\cite{Scheel2008}.  The primary disadvantage of these
methods is that they require fine tuning during the plunge
(Sec.~\ref{sec:Plunge}). For example, the function $g(x,t)$ defined in
Eq.~(\ref{eq:GaugeRolloff2}) must be chosen carefully or else the
simulation fails shortly (a few $M$) before a common horizon forms.
There are at least two reasons that fine tuning is currently
necessary. First, the gauge conditions must be chosen so that no
coordinate singularities occur before merger. Second, the excision
boundaries do not coincide with the apparent horizons, but instead
they lie somewhat inside the horizons.  If the excision boundaries
exactly followed the horizons, then the characteristic fields of
the system would be guaranteed to be outflowing (into the holes) at
the excision boundaries, so that no boundary condition is required
there.  But for excision boundaries inside the horizons, the outflow
condition depends on the location of the excision boundary, its motion
with respect to the horizon, and the gauge.  Indeed, the most common
mode of failure for improperly-tuned gauge parameters is that the
outflow condition fails at some point on one of the excision
boundaries.  We have been working on improved gauge
conditions~\cite{Lindblom2009c} and on improved algorithms for
allowing the excision boundary to more closely track the apparent
horizon.  These and other improvements greatly reduce the amount of
necessary fine tuning and allow mergers in generic configurations, and
will be described in detail elsewhere~\cite{Szilagyi2009}.

\begin{figure}
\includegraphics[width=0.47\textwidth]{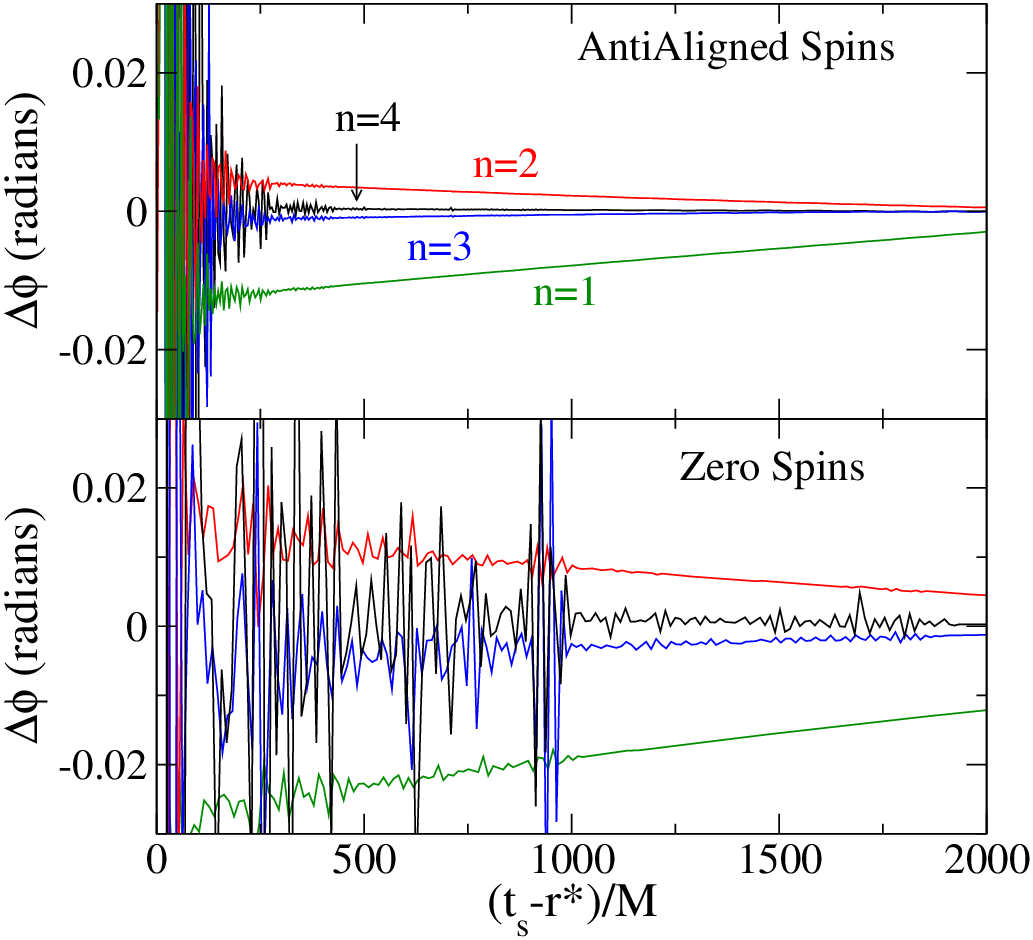}
\caption{\label{fig:0093c_vs_DD}
Comparison of waveform
extrapolation between the current simulation of counter-rotating
black holes (top panel), and the earlier simulation of
non-spinning black holes~\cite{Boyle2007,Scheel2008}.  
The noise is significantly reduced in
the newer simulation, due to smaller constraint damping parameters
in the wave zone.
}
\end{figure}

Another quite important improvement lies in the choice of
  constraint damping parameters.  To illustrate this effect, 
Fig.~\ref{fig:0093c_vs_DD} compares the gravitational wave phase 
extrapolation for the simulation presented here with the similar plot for 
an earlier run~\cite{Boyle2007} with different constraint damping 
parameters.  As can be seen in
Fig.~\ref{fig:0093c_vs_DD}, the improved constraint damping
parameters result in significantly reduced noise.  For the
earlier simulation, the waveform was unusable for $t-r^*<1000M$, 
and was still noticeably noisy at $1000M<t-r^*<2000M$. For the new simulation,
  the smaller constraint damping parameters
  result in clean waveforms as early as $t-r^*\sim 250M$, despite the
  observation that the spinning black holes
  result in a pulse of junk radiation of about twice the amplitude 
  of the earlier run.  
  The new simulation also shows smaller
  extrapolation errors, presumably because the new simulation uses
  larger extraction radii (up to $r=350M$, whereas
  Ref.\cite{Boyle2007} uses a largest extraction radius of $r=240M$).

We employ four techniques to measure black hole spin: Two of these are
based on the surface integral for quasi-local linear momentum, and
utilize either simple coordinate rotation vectors $\chi_{\rm Coord} (t)$
or approximate Killing vectors, $\chi_{\rm AKV} (t)$; the other two are
based on the shape of the apparent horizon, and infer the spin from
the extrema of the scalar curvature ($\SpinFromShapeMin (t)$,
$\SpinFromShapeMax (t)$).  The four spin measures agree to better than 1
per cent during the inspiral.  The AKV spin $\chi_{\rm AKV} (t)$ shows the
least variations during the simulation, and is the only
spin diagnostic that results in a monotonically decreasing spin during
the inspiral, as expected from the effects of tidal friction.  The
other three spin measures ($\chi_{\rm coord} (t)$, $\SpinFromShapeMax (t)$,
$\SpinFromShapeMin (t)$) show various undesired and physically
unreasonable behaviors: All three result in {\em increasing} spin
during the inspiral, inconsistent with tidal friction
(cf. Fig.~\ref{fig:AntiAlignedInspiralSpins}).  $\SpinFromShapeMin (t)$
and $\SpinFromShapeMax (t)$, furthermore show very strong variations
during the initial transients, just before merger, and just after the
common horizon forms.  This is expected, as in those regions of the
evolution, the black holes can not be approximated as isolated Kerr
black holes.  The behavior of $\SpinFromShapeMin (t)$ and
$\SpinFromShapeMax (t)$ contain information about the deformation of the
black holes.  The final state of the simulation is expected to be a
single, stationary Kerr black hole, for which $\SpinFromShapeMin (t)$ and
$\SpinFromShapeMax (t)$ should result in the correct spin.  Indeed, all
four spin diagnostics agree at very late time to five significant
digits (cf. Fig.~\ref{fig:AntiAlignedSpinsRingdown}).

\begin{table}
\begin{tabular}{|lcc|}
\hline
Prediction Formula & $\chi_f$ & $M_f/M$\\ 
\hline
Kesden~\cite{Kesden2008}&  0.521153   & 0.97039\\
Buonanno, Kidder \& Lehner~\cite{Buonanno2008a}  & 0.505148   & 1.0\\
Tichy \& Marronetti~\cite{Tichy2008} &   0.548602 & 0.962877\\
Boyle \& Kesden~\cite{Boyle2007b} & 0.547562     & 0.964034 \\
Barausse \& Rezzolla~\cite{Barausse2009} & 0.546787  & 1.0\\
\hline
Numerical result (this paper)   & 0.54781 & 0.961109  \\ 
\hline
\end{tabular}
\caption{
Predictions of final black hole
spin and mass from analytical formulae in the literature,
applied to the simulation considered here. 
Refs.~\cite{Buonanno2008a,Barausse2009} do not predict 
the final mass, but instead assume zero mass loss.
\label{tab:PredictionTable}
}
\end{table}

The accuracy of our simulation places new constraints on analytic
formulae that predict the final black hole spin from the initial spins
and masses of a black hole binary.  Table~\ref{tab:PredictionTable}
lists some of these predictions.

\acknowledgments{
We would like to thank Lee Lindblom, B\'ela Szil\'agyi, and Kip Thorne
for helpful discussions and comments.
We thank Mike Kesden for computing the final mass and
spin predictions from the
various analytical models.
We are especially grateful to
Fan Zhang for computing the extrapolation of the waveforms to infinite
radius using methods and a variant of code developed by Mike Boyle 
and Abdul Mroue.
This work was supported in part by the Sherman Fairchild Foundation, 
the Brinson Foundation, by NSF grants PHY-0601459, PHY-0652995, and DMS-0553302
and by NASA grant NNX09AF97G.
HP acknowledges support from the Canadian Institute for Advanced Research.
Some calculations were done on the Tungsten cluster at NCSA. 
}


\end{document}